\documentclass[12 pt]{article}

\usepackage{setspace,relsize}

\usepackage{moreverb}

\usepackage[margin=1.3in]{geometry}

\usepackage[pdftex]{lscape}
\usepackage{amscd}
\usepackage{amsmath}
\usepackage{graphicx}
\usepackage{amsfonts}
\usepackage{amssymb}
\usepackage{setspace}
\usepackage{natbib}
\usepackage{mathrsfs}
\usepackage{multicol}
\usepackage{multirow}
\usepackage{pifont}
\usepackage{enumerate}
\usepackage{epic}
\usepackage{color}
\usepackage{booktabs}
\usepackage{authblk}
\usepackage{algorithm}
\usepackage{algpseudocode}
\usepackage{epstopdf}
\usepackage{subfig} 
\usepackage{newfloat} 
\usepackage{adjustbox} 

\DeclareMathOperator*{\argmax}{arg\,max} 

\title{\large Differentially Expressed Functional Connectivity Networks with K-partite Graph Topology}

\author{\small Shuo Chen$^1$\thanks{Correspondence to: shuochen@umd.edu},  F. DuBois Bowman$^2$, Yishi Xing$^1$,    \\
\small $^1$ Department of Epidemiology and Biostatistics, University of Maryland, College Park, MD 20742, USA \\

\small $^2$  Department. of Biostatistics, Columbia University

}
\date{}

\begin{document}
\doublespacing \maketitle
 
\begin{abstract}
Emerging brain network studies suggest that interactions between various
distributed neuronal populations may be characterized by an organized complex topological structure. Many brain diseases are associated with altered topological patterns of brain connectivity. Therefore, a key inquiry of  connectivity analysis is to identify network-level differentially expressed connections that have low false positive rates, sufficient statistical power,
and high reproducibility. In this paper, we propose a novel statistical approach to fulfill this goal by leveraging the topological structure of differentially expressed functional connections or edges in a graphical representation. We propose a new algorithm to automatically  detect the latent topology of a k-partite graph structure, and we also provide statistical inferential techniques to test the detected topology. We evaluate our new methods via extensive numerical studies. We also apply our new approach to resting state fMRI data (24 cases and 18 controls) for Parkinson's disease research. The detected connectivity network biomaker with the k-partite graph topological structure reveals underlying neural features distinguishing Parkinson's disease patients from healthy control subjects.

\end{abstract}

\emph{Keywords}: brain connectivity, fMRI, graph topology, k-partite, differentially expressed networks,    Parkinson's disease.


\section{Introduction}
Neuroimaging research has suggested that neuropsychiatric disorders are associated with altered interactions between distributed neuronal populations and brain regions (\citealp{Buckner09}; \citealp{Craddock13}; \citealp{Stam14}; \citealp{Fornito15}; \citealp{Chen15a}). Recent neuroimaging studies have used graph theory as a tool to understand the brain connectivity patterns, which denote brain regions as nodes, and connections between them as edges (\citealp{Bullmore09}; \citealp{Robinov10}; \citealp{Craddock13}; \citealp{Biswal10}; \citealp{Yeo11};   \citealp{Sporns14}; \citealp{Smith15}).  Such studies have identified  connectome-phenotype relationships by leveraging these graph techniques, mainly using  network graph descriptive metrics (\citealp{Bullmore09};  \citealp{Seeley09}; \citealp{Stam09}; \citealp{Robinov10}; \citealp{Achard12}; \citealp{vand13}; \citealp{Crossley13}; \citealp{Crossley14}; \citealp{Stam14}; \citealp{Fornito15}). 

The overarching goal of brain connectivity network/circuitry  research is to enhance understanding of underlying pathophysiological mechanisms and clinically useful predictions concerning disease diagnosis and treatment selection (\citealp{Fornito12}; \citealp{Craddock13}; \citealp{Fornito15}). However, generally it is challenging to detect such differential connectivity networks  that simultaneously i) control false positive rate and obtain sufficient statistical power; ii) reflect complex connectome topological properties; iii) are spatially localized (edge -specific) for explicit clinical interpretation and pathophysiological mechanism discovery; and iv)  are reliable and reproducible (\citealp{van13}; \citealp{Simpson15}). A driver of these challenges is the nature of connectome data, containing  complex  topological structure and high-dimensionality. 
Most network graph  metrics summarize all edges as individual measures and  lose localized connectivity information (edge - specific). Such measures may lead to difficult clinical interpretability and may lack specificity and sensitivity (\citealp{Simpson15}). On the other hand, the mass univariate analyses may keep localized information but are subject to the trade-off between false positive discovery control and lack of statistical power. Direct application of the family-wise error control and false positive discovery rate control methods could successfully prohibit spurious positive findings, yet they may be overly conservative and reduce the statistical power  and lead to few or no significant findings. To mitigate such trade-offs, many studies pre-define networks of interest to lower the multiple testing burden. But, pre-defined networks are limited and may exclude   potential true signals. Recently, more advanced statistical methods leverage multivariate models to link the edge connection strength and overall topological structure to improve model estimation (\citealp{Simpson12}; \citealp{Simpson15b}). However, the resulting inferences (regression coefficients) regarding phenotype-related connectome features by using these methods are edge-specific rather than at the network level, and thus they may not allow to automatically detect differentially expressed subgraphs.
 
In this paper, we define a differentially expressed brain connectivity network as an object that includes three components: nodes (brain areas), edges (connections between brain areas) that are differentially expressed between clinical groups, and topological structures of the graph consisting of these nodes and edges. The topological structure is automatically detected from the data rather than pre-specified. Differentially expressed edges may be distributed in an organized pattern rather than randomly in the brain. Therefore, we  consider the union of  differentially expressed edges and their organized topological pattern as a potential brain connectivity subnetork biomarker candidate.   A differentially expressed subnetwork  can increase statistical power while effectively controlling multiple testing false positive errors by allowing edges borrow strength with each other within the topological structure. Statistical methods have been developed to test the statistical significance of a subset of edges, for example, network based statistic (NBS) and parsimonious differential network detection (Pard) (\citealp{Zalesky10}; \citealp{Chen15b}). The Pard algorithm seems to be more powerful as it not only recognizes the differentially expressed edges but also the topology structure of these edges by applying the rule of parsimony. Hence, automatic detection of latent topological structure  is the key step for potential connectivity network biomarker detection. The more accurately we can identify the underlying topological structure, the lower false positive and negative error rates can be achieved.  In addition, the detected topological structure can provide assistance to reveal the underlying neurophysiologcial and neuropathological mechanism for brain science research. For example, the clique structure of a differential network suggests most edges between the  nodes of the network are differentially expressed, prompting further examination of the specific brain regions involved and their associated interactions. The Pard algorithm only detects the complete subgraph structure of a differential network, however more complex topological structure may appear.

In this article, we propose a novel statistical strategy to detect an organized graph topology: a k-partite  graph. In graph theory, a k-partite subgraph is a graph whose nodes could be partitioned into k distinct sets such that the nodes in the same set are not connected but nodes from different sets  are connected. For brain connectivity analysis, an edge is often a continuous (rather than binary) quantity that represents i) the connection strength for one subject (e.g. the Pearson correlation coefficient) or ii) to what extent the connection is differentially expressed among different clinical groups (e.g. a test statistic). In this paper, as we try to identify the differentially expressed subnetworks we use the later case. Specifically, we refer to ``k-partite phenomenon" (for illustration we let k=2) when: i)  there are two sets of nodes and the nodes within the same set are highly connected  for all subjects across clinical subpopulations; ii) the within set connections  are not expressed differentially between clinical groups; iii) the between set connections show difference across clinical groups. If we observe ``k-partite phenomenon", we may infer that the brain disease alters the   long-range  interactions between two sets of neural populations (nodes) rather than the local interactions. If we successfully recognize the topological structure of the differential subgraph, we greatly improve the statistical power and effectively control false positive findings simultaneously.  However, in practice the ``k-partite phenomenon" is latent and not directly observable. Therefore, we develop a novel statistical framework to detect the structure and provide statistical inference methods to test the statistical significance  of the topological structure.

\section{Methods}

We consider data from S subjects, potentially representing distinct subgroups (e.g. based on the presence of a clinical diagnosis).  fMRI data from each subject undergoes preprocessing and is used to calculate functional connectivity between $n$ nodes or brain regions, with whole-brain connectivity for a subject $s$ represented  by an $n\times n$ symmetric matrix $\mathbf{M}^s$.  Therefore, the overall data set is denoted by $\mathbb{M}= \{\textbf{M}^1, \cdots, \textbf{M}^S \}$.  The location (3D coordinates) of a node $i$ ($i \in 1, \cdots, n$)  is identical for all subjects.  We denote $M^s_{ij}$ as the connectivity metric between node $i$ and node $j$ for subject $s$. For instance the functional connectivity (FC) edge could be calculated by using correlation coefficient (or other metrics such as mutual information coefficients) between two time courses from two nodes using fMRI data. 

Next, we perform statistical analysis (e.g. two sample test or regression for a case-control study) on each edge and record a p-value $p_{ij}$ between nodes $i$ and $j$. The direct inference on $\{ p_{ij} \}$ could be invalid due to the dependence structure between edges and the multiple testing issue. Instead of making inference by $p_{ij}$, we use a weighted matrix \textbf{W} with the entry of $i$th row and $j$th column $W_{ij}=-\log(p_{ij})$ as our input weighted adjacency matrix. We utilize $W_{ij}$ as a metric to delineate the informative level (how differentially expressed between clinical groups) of an edge. For instance, \textbf{W} is the input matrix for both NBS and Pard algorithms, and the outputs of these algorithms are subsets of edges with significance levels $\{ G_c \}$ ($c=1, \cdots, C$ and C is the number of significant structures). Usually, the Pard will identify $G_c$ as a clique/block with high proportion of informative edges (see  Figure \ref{fig3:f1} and \ref{fig3:f2} for Pard clique subgraph detection as an example). In brief, the Pard algorithm tries to include as many  informative edges as possible in the minimal network (block) and tests whether there are network differences between clinical groups via permutation test (\citealp{Chen15b}). In this article, we develop our methods based on a detected network $G_c$ (with corresponding $\textbf{W}_c$ see Figure \ref{fig3:f3} for example) by NBS or Pard, and $\textbf{W}_c$ is considered as the input data of our new  \textbf{k}-\textbf{p}artite \textbf{g}raph \textbf{d}etection (KPGD) algorithm.

\subsection{k-partite graph detection}

\begin{figure}[!tbp]
  \centering
  \subfloat[An implicit bipartite graph]{\includegraphics[width=0.5\textwidth]{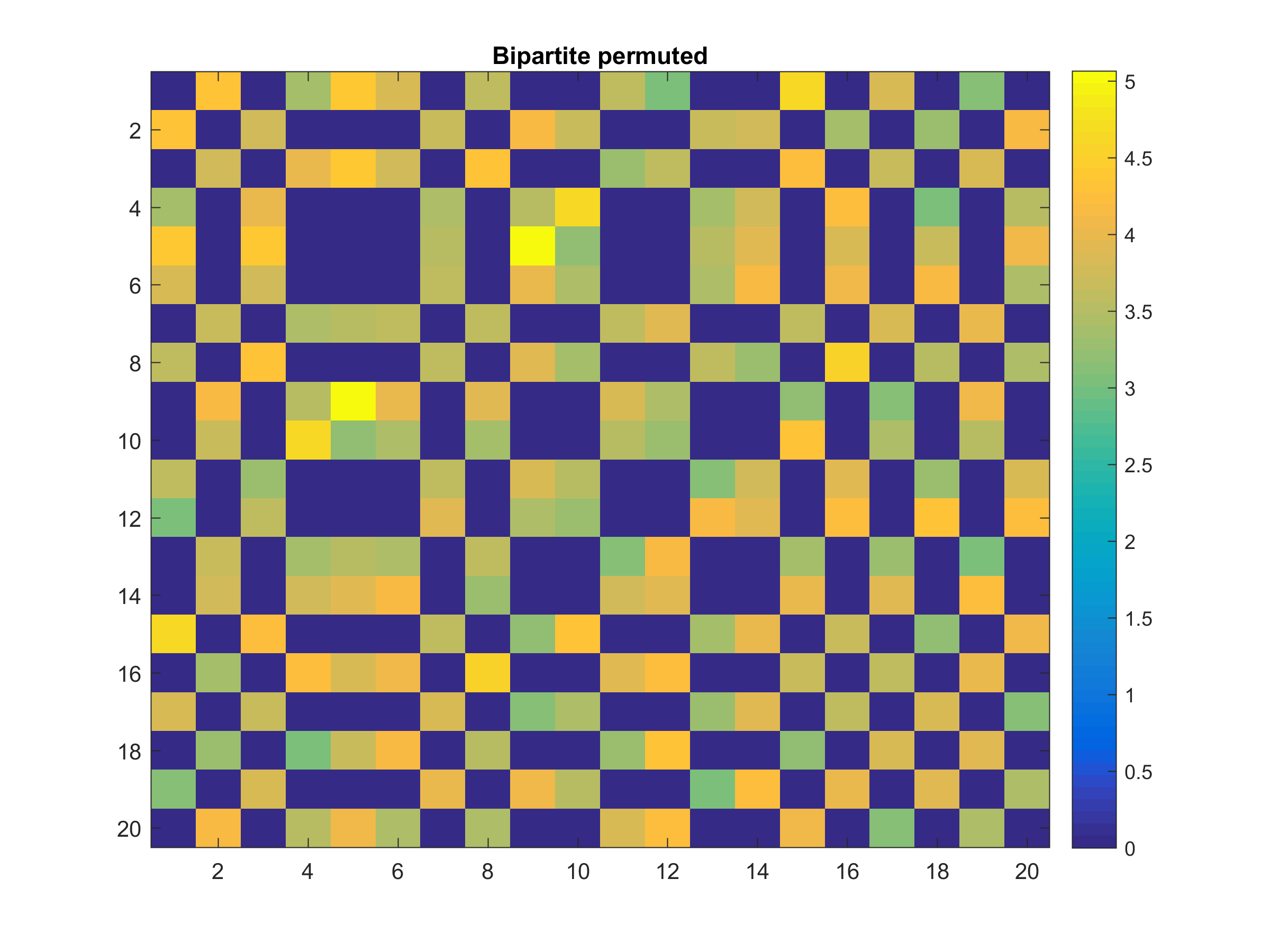}\label{fig1:f1}}
  \hfill
  \subfloat[Detected bipartite graph]{\includegraphics[width=0.5\textwidth]{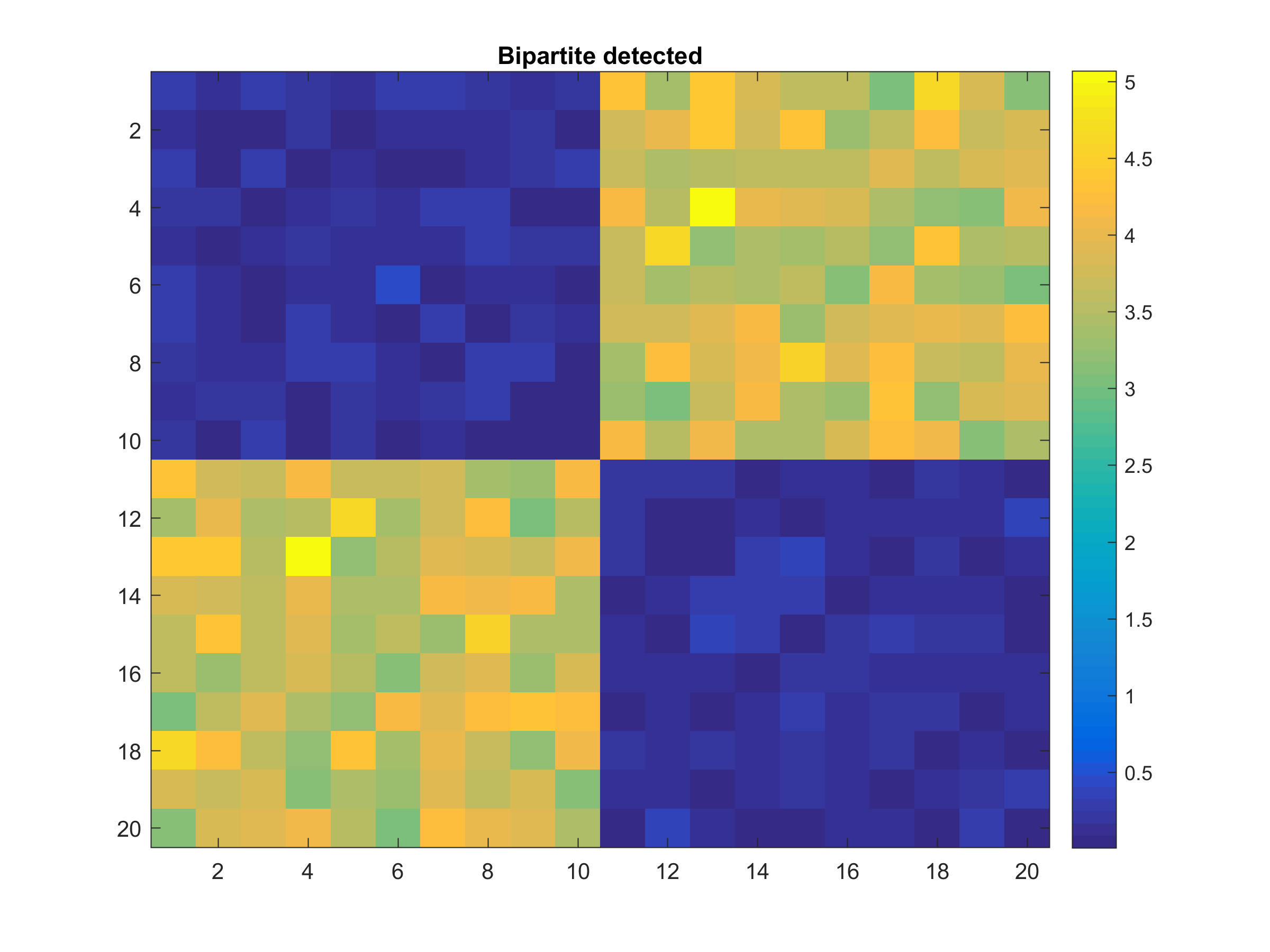}\label{fig1:f2}}
  \caption{An example of a k-partite (k=2) graph  $G_c$  with  20 nodes: (a) is the k-partite graph $G_c$ with the original order of nodes that is comparable to our input data with a latent k-partite topological structure; and (b) is output graph after applying KPGD algorithm that is an isomorphic graph of $G_c$ with reordered nodes and an explicit topological structure.}
\end{figure}

We let $G_c$  be a k-partite graph. Note that, as illustrated in Figure \ref{fig1:f1}, the k-partite pattern of $G_c= \{V_c, E_c\}$ is latent. 
To observe the k-partite graph explicitly as shown in Figure \ref{fig1:f2}, we need to reorder the nodes. In our algorithm,  we seek an edge preserving bijective mapping (node reordering) function $\pi$ that  $\pi: G_c \rightarrow H_c$ or $H_c=\pi(G_c) $. In graph theory, $G_c$ is isomorphic to $H_c$ ($G_c \simeq H_c$), where $H_c= \{V'_c, E'_c \}$. The bijective mapping functions $\pi$ permutes the order of nodes (in columns and rows of the connectivity matrix) simultaneously. If nodes $i$ and $j$ are connected, denoted $i \sim j$, then after permutation $\pi(i) \sim \pi(j)$. Figure \ref{fig1:f2} shows a desirable mapping function $\pi$ because in $H_c=\pi(G_c) $ the bipartite structure is directly apparent. However, the number of all possible permutations is massive with $n!$ permutations   for a graph with $n$ nodes. For example, when $n$=100  there are more than $10 \times 10^{157}$ possible permutations. Therefore, it is impractical to exhaust all permutations, and we need an algorithm to seek an appropriate mapping $\pi$ that reveals the k-partite structure. Note that edges in our input matrix  $\textbf{W}_c$ are weighted. The key heuristic of our algorithm is that the target mapping function $\pi$ allocates more informative edges (small $p_{ij}$ values and more significant connectivity difference between clinical groups) to off-diagonal blocks and less informative edges along the diagonal blocks. The diagonal blocks represent independent sets of a k-partite graph, where edges within the same independent set are not connected (not informative for our case).

We define the objective function of the KPGD algorithm as:

\begin{equation} 
 \argmax_{\{ A_k^c\}_{k=1}^{K_c}} \sum_{k=1}^{K_c} \frac{\sum_{i \in   A_k^c, j \not\in  A_k^c}-\log(p_{ij})}{ | A_k^c|} ,
\label{fm1}
\end{equation}

where Formula \ref{fm1} maximizes the informative edges between the $K_c$ independent sets with a given number $K_c$, and $A_k^c$ represents an independent set. The denominator $| A_k^c|$ equals to the number of nodes in $A_k^c$ and is used to avoid a singleton set. However, the direct optimization of the objective function  \ref{fm1} is nondeterministic polynomial time (NP) complex  (\citealp{von07}).  

We consider an approximate solution via discretized relaxation by using spectral graph theory. Let $\mathbf{X}^c_{|V_c| \times K_c} \in \{0,1 \}$ be a binary group index matrix with $|V_c|$ rows (number of nodes in $G_c$) and $K_c$ columns. $\mathbf{X}_{|V_c| \times K_c}  \textbf{1}_{K_c \times 1}=\textbf{1}_{|V_c| \times 1}$ indicates that a node only belongs to one independent set exclusively. Thus, our k-partite detection algorithm is equivalent to estimate $\hat{X^c}$ by maximizing the objective function in formula \ref{fm1}, and we can rewrite it as: 

\begin{equation} 
 \argmax_{X^c} \sum_{k=1}^{K_c} \frac{\mathbf{x}_k  \mathbf{L}^c x_k  }{ \mathbf{x}_k^{ '} x_k }, 
\label{fm2}
\end{equation}
where $\mathbf{L}^c=\mathbf{D}^c-\textbf{W}_c$  and $\mathbf{D}$=diag \{ $\sum_{j=1}^{|V_c|} W_{1j}, \cdots, \sum_{j=1}^{|V_c|} W_{|V_c|j} \}$ . $\mathbf{L}^c$ is the Laplacian matrix (that is semi-positive definite), and $\mathbf{x}_k$ is a vector: the $k$th column of $\mathbf{L}^c$.  The optimization is Raleigh quotient when $\mathbf{x}_k$ is a continuous vector, and bounded by the sum of the largest $k$ eigen values of $\mathbf{L}$ (\citealp{von07}, \citealp{Horn12}). Similar to the discretization relaxation of spectral clustering algorithms, we next perform K-means clustering algorithms on the  largest  $K_c$ vectors to estimate $\widehat{\mathbf{X}^c} $ (the membership of the independent sets). Note that the KPGD algorithm performs K-means clustering on the  largest,  rather than the smallest,  $K_c$ vectors of the Laplacian matrix because our goal is to allocate all informative edges to off-diagonal blocks.

Clearly, the selection of $K$ is crucial for k-partite structure detection.  
We apply a data-driven and automatic $K$ selection by using the `quantity and quality' criteria that ensures i) all informative edges are moved to the off-diagonal (`quantity') and the proportion of informative edges in the off-diagonal blocks are high -- concentrated connections between independent sets (`quality') (\citealp{Chen15b}).
\begin{equation} 
\frac{\sum_{k=1}^{K_q} \sum_{i \in A_k^c, j  \notin A_k^c} I(W_{ij}>0)}{\sum_{i,j \in G_c} I(W_{ij}>0)} \cdot \frac{\sum_{k=1}^{K_q} \sum_{i \in A_k^c, j  \notin A_k^c} I(W_{ij}>0)}{\sum_{k=1}^{K_q} \sum_{i \in A_k^c, j  \in A_k^c} 1},
\label{fm3}
\end{equation}

where is the first term is the `quantity' criterion,  and the second term is the `quality' criterion. \textbf{W} is thresholded according to: 

\[ W_{ij} = \left\{ \begin{array}{ll}
         -\log(p_{ij}) & \mbox{if $ p_{ij} \leq p_0$};\\
        0 & \mbox{if $p_{ij} > p_0$},

        \end{array} \right. \]
 to avoid the accumulation of noise-related false positives..  Therefore, we select the optimal $k^c$ by grid searching (from $2, \cdots, K^c$).  We summarize the whole procedure in Algorithm \ref{KPGD}.     
        
\begin{algorithm}
\caption{KPGD algorithm}
\label{KPGD}
\begin{algorithmic}[1]
\Procedure{KPGD\textendash Algorithm}{}
 
\State Calculate the Laplacian matrix $L^c=D^c-\textbf{W}_c$;
\For{cluster number  $K_c$ = 2 to $|V_c|-1$ }
\State Compute the first $K_c$ eigenvectors $[u_1, \cdots, u_{K}]$ of $L^c$, with eigenvalues ranked from the largest;
\State Let $U={[u_1^T, \cdots, u_{K}^T]}$ be a $|V_c|\times K_c$ matrix containing all $K_c$ eigenvectors;
\State Perform K-means clustering algorithm on $U$ with $K$ to cluster $|V_c|$ nodes into $K_c$ independent sets;
\State Calculate the quality and quantity criterion for each $K_c$.
\EndFor
\State Adopt the clustering results using the $K_c$ of the maximum score of the quality and quantity criterion by formula \ref{fm3}.
 
\EndProcedure
\end{algorithmic}
\end{algorithm}

\subsection{Statistical test for k-partite structure}

In section 2.1, by applying the KPGD algorithm, we detect the independent sets $\{ A_k^c \}$ from $G_c$, where the intra-set edges are less informative than inter-set edges. In this section, we propose a statistical test to verify the k-partite structure (i.e. the organized pattern of informative edges) is genuine and then provide a p-value for statistical inference. Since  $G_c$ has been detected as a differentially expressed subgraph by using algorithms such as NBS or Pard, our goal is to further test whether the informative edges of $G_c$ are distributed in a k-partite structure $H_c=\pi(G_c)=\cup_{k=1}^{K^c} A_k^c$ or just randomly. Therefore, our null and alternative hypotheses are:

$H_0$: The informative edges  are distributed randomly in $G_c$.

$H_1$: The informative edges  are distributed as a k-partite pattern in $G_c$.

We apply the `graph edge permutation' test to determine the statistical significance of the non-randomness of the k-partite topological structure (Chen et. al 2016). The non-parametric permutation test for statistical inferences is appealing here, given the challenges posed for determining appropriate asymptotic distributions of the test statistic based on the complex object $G_c$ and multiple testing issues (\citealp{Nichols02}; \citealp{Zalesky10}; \citealp{Winkler14}). 



The edge permutation test shuffles the order/location of each edge in $G_c$. We first transform the input matrix $\textbf{W}_c$ of $G_c$ into a vector of edges $E_c$ with the length of $|V_c| \times (|V_c|-1)/2$ ($|V_c|$ is the number of nodes in  $G_c$), that $E_c= \{W_{1,2}^c, W_{1,3}^c,\cdots,  W_{|V_c|-1,|V_c|}^c \}$. The edge permutation can be considered as a mapping function $\phi$ that projects $\textbf{W}_c$ to $\textbf{W}^m_c=\phi(\textbf{W}_c)$ for the $m$th permutation.  But, the mapping is not edge preserving as $\textbf{W}_c(i,j) \neq \textbf{W}^m(i,j)$. Therefore, suppose $G_c$ is a subgraph with a specific non-random topological structure e.g. k-partite,  and $\phi (G_c)$ is likely to be a random graph.  Here, we use a test statistic $T_c^0= 1/|E_c^{off}| \sum_{k=1}^{K_c}  {\sum_{i \in   A_k^c, j \not\in  A_k^c}-\log(p_{ij})}  -   1/|E_c^{diag}| \sum_{k=1}^{K_c}  {\sum_{i \in   A_k^c, j \in  A_k^c}-\log(p_{ij})}$ to contrast the differential levels of edges between the independent sets (off-diagonal blocks) and within theses sets (diagonal blocks). Since the proportion of informative edges in $G_c$ is high as a selected subgraph, if the informative edges are distributed in an organized pattern, the test statistic of $\textbf{W}_c$ should be greater than those calculated for most of the permutations $\textbf{W}^m_c$ ($m=1, \cdots, M$). Therefore, we leverage the edge permutation to examine the hypotheses $H_0$ and $H_1$ regarding the topological pattern of $G_c$, and the detailed  algorithm is described in Algorithm \ref{GEP}. 

\begin{algorithm}
\caption{Graph Edge Permutation (GEP) Test}
\label{GEP}
\begin{algorithmic}[1]
\Procedure{GEP \textendash Algorithm}{}

\State  We apply the KPGD algorithm to $\textbf{W}_c$, and calculate a statistic $T_c^0= 1/|E_c^{off}| \sum_{k=1}^{K_c}  {\sum_{i \in   A_k^c, j \not\in  A_k^c}-\log(p_{ij})}  -   1/|E_c^{diag}| \sum_{k=1}^{K_c}  {\sum_{i \in   A_k^c, j \in  A_k^c}-\log(p_{ij})}$, where $|E_c^{off}|$ is the number of edges between the independent sets and $|E_c^{diag}|$ is the number of edges within the independent sets.

\State List the all edges in $G_c$ as a vector in the original order, $ vec(\textbf{W}_c)= \{W^c_{1,2} , \cdots, W^c_{|V_c|-1, |V_c|} \}$, $|V_c|$ is the number of nodes and $W^c_{i,j}=- \log (p_{ij})$. 
 
\For{each permutation iteration  $m =1:M$ }

\State Shuffle the order of edges in $vec(\textbf{W}_c)$, and obtain an edge reordered graph $G_c^m$ with a weighted edge matrix $\textbf{W}_c^m$;
\State Apply the KPGD algorithm on $G_c^m$ (or $\textbf{W}_c^m$) and obtain $K_c^m$ independent sets, where $K_c^m$ is determined by the `quantity and quality' criteria;
\State Calculate the test statistic $T_c^m$ as described in line 2.   
\EndFor

\State If $T_c^0$ is greater than the top 5th  percentile of $T_c^m$, we reject the Null Hypothesis, and thus $G_c$ has a k-partite topological structure . 

\EndProcedure
\end{algorithmic}
\end{algorithm}

In summary, we perform KPGD for pattern recognition (object oriented statistical estimation) and GEP for pattern significance test (object oriented statistical inferences).  The joint application of the KPGD algorithm and the permutation test provides a pathway to identify a differentially expressed subgraph with a k-partite graph topology. The differentially expressed brain connectivity network with a k-partite graph topology is more specific because it is a part of $G_c$ (without diagonal blocks). Therefore, the informative edges could borrow power from each other more efficiently within a subgraph where most informative edges concentrate. If the k-partite graph topology is true, our method will not only effectively increase statistical power to detect the  differentially expressed edges with low false positive discovery rate but also provide a topology to describe the topological pattern of these  differentially expressed edges.

\section{Numerical Results}
We conduct simulation studies to evaluate whether the KPGD algorithm can optimally determines the number of independent sets and recognize the independent sets in different settings (i.e. the k-partite structure). 

We first generate a $20 \times 20$ $\textbf{W}_c$ matrix with each edge given by $-\log(p_{ij})$. Next, we let the number of independent components (k) of the k-partite graph equals to 2, 4, and 10, and we use larger values of $-\log(p_{ij})$ for off-diagonal edges. For example, we let the edges within the diagonal blocks to be around $-\log(0.35)=1.05$, and edges within the off-diagonal blocks to be around $-\log(0.02)=4.0$. We set the center difference between diagonal and off-diagonal block edges as a parameter $\delta$.  We are unaware of alternative algorithms to automatically detect k-partite topological structure based on weighted-edge graph, and thus we do not include the performance of competing methods. 

We permute the order of the nodes, and the k-partite structures are not directly observable (subfigures of the left column, Figure 2). We apply the KPGD algorithm to detect the topological structure and the GEP test for statistical inferences. For all settings, our algorithm successfully identifies the correct numbers of independent sets $k$ and reveals the k-partite graph. 

We evaluate the performance of our new methods by simulating 100 data sets with different values of both   $k$  and   $\delta$. We summarize our results in table 1, specifically reporting rates of false positive (FP) and false negative (FN) edges. We observe that in general the KPGD algorithm performs well, and the GEP testing strategy successfully recognizes k-partite structure.  When $k$ is smaller, the FP and FN is lower, because the pattern is closer to a complete subgraph (clique) when $k$ is large. Moreover, our `quantity and quality' rule successfully determines the optimal $k$, and we consistently estimate the true $k$ correctly for each simulated data set (more than 99\%). The GEP algorithm  also provides  sufficient statistical power to reject the null hypothesis that the subgraph $G_c$ is random.  Also, our methods seem to be robust to the choice of $\delta$, and hence the detected topological pattern could improve the  robustness to mild and moderate increase of the noise to signal ratio.  

\begin{figure}[!htp]
  \centering
  \subfloat[Input bipartite $\textbf{W}_c$]{\includegraphics[width=0.5\textwidth]{2partiteperm.png}\label{fig2:f1}}
  \hfill
  \subfloat[Detected bipartite graph by the KPGD algorithm]{\includegraphics[width=0.5\textwidth]{2partitedetec.png}\label{fig2:f2}}
  \hfill
  \subfloat[Input 4-partite $\textbf{W}_c$]{\includegraphics[width=0.5\textwidth]{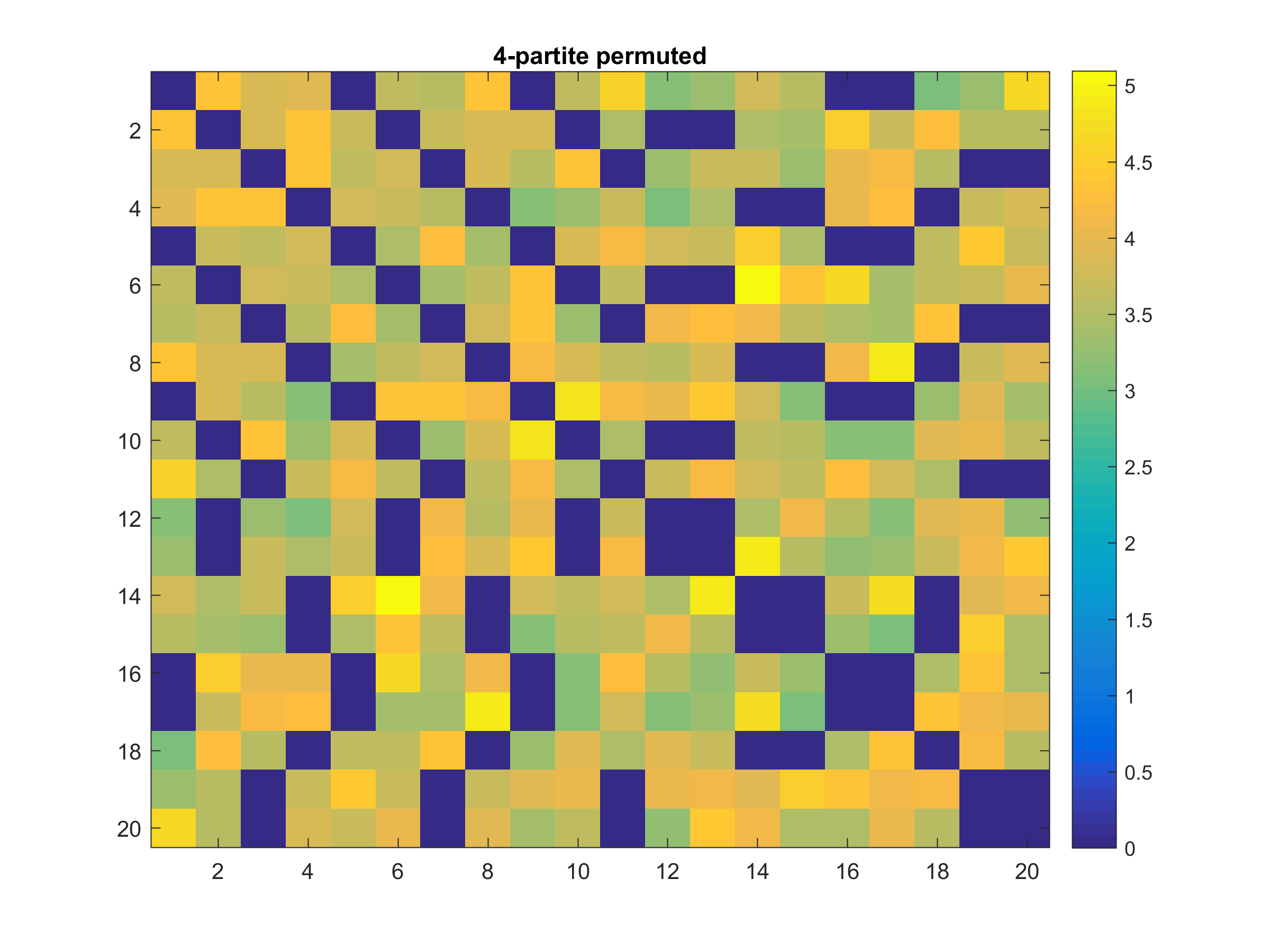}\label{fig2:f3}}
  \hfill
  \subfloat[Detected 4-partite graph by the KPGD algorithm]{\includegraphics[width=0.5\textwidth]{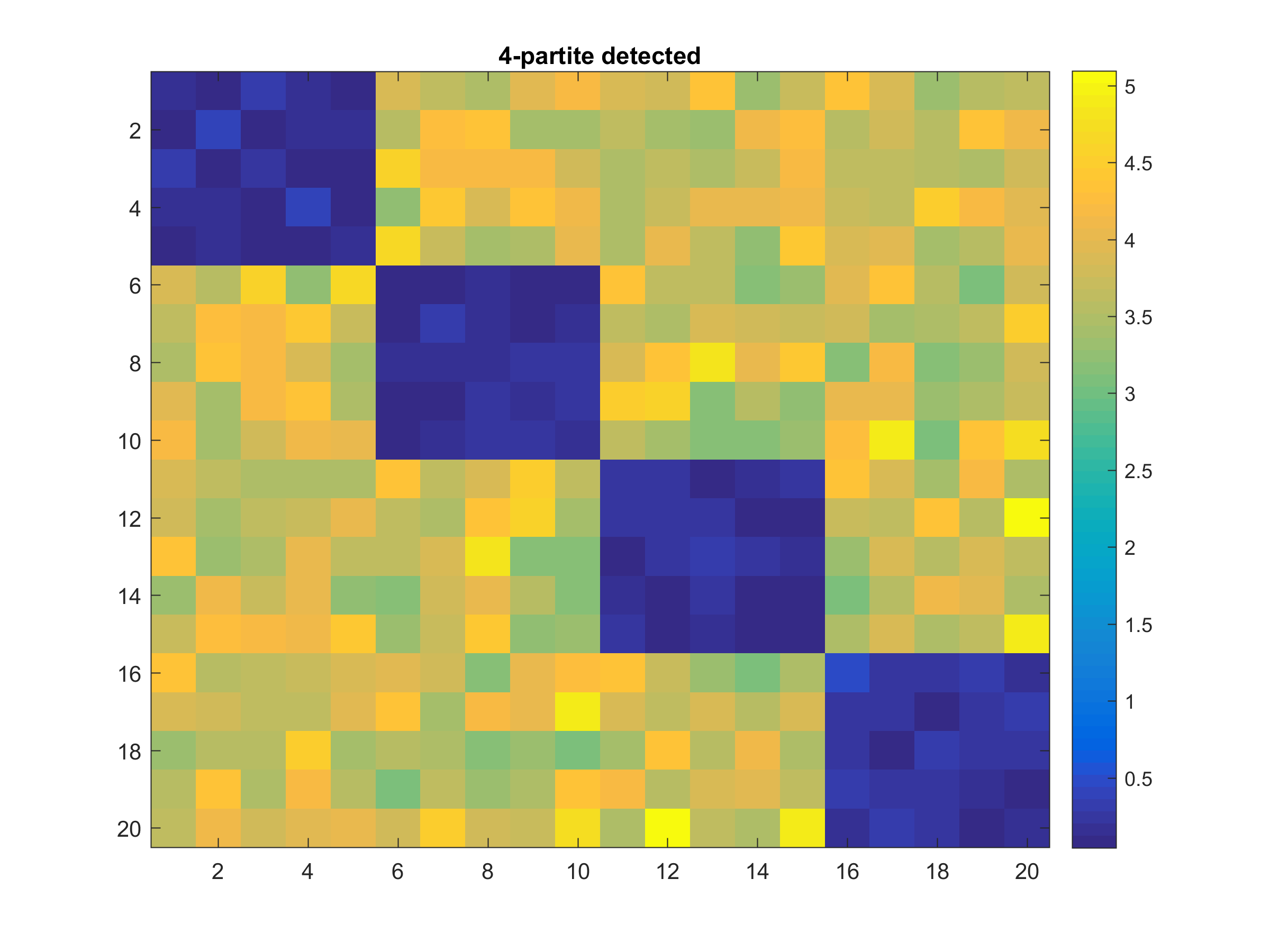}\label{fig2:f4}}
  \hfill
  \subfloat[Input 10-partite $\textbf{W}_c$]{\includegraphics[width=0.5\textwidth]{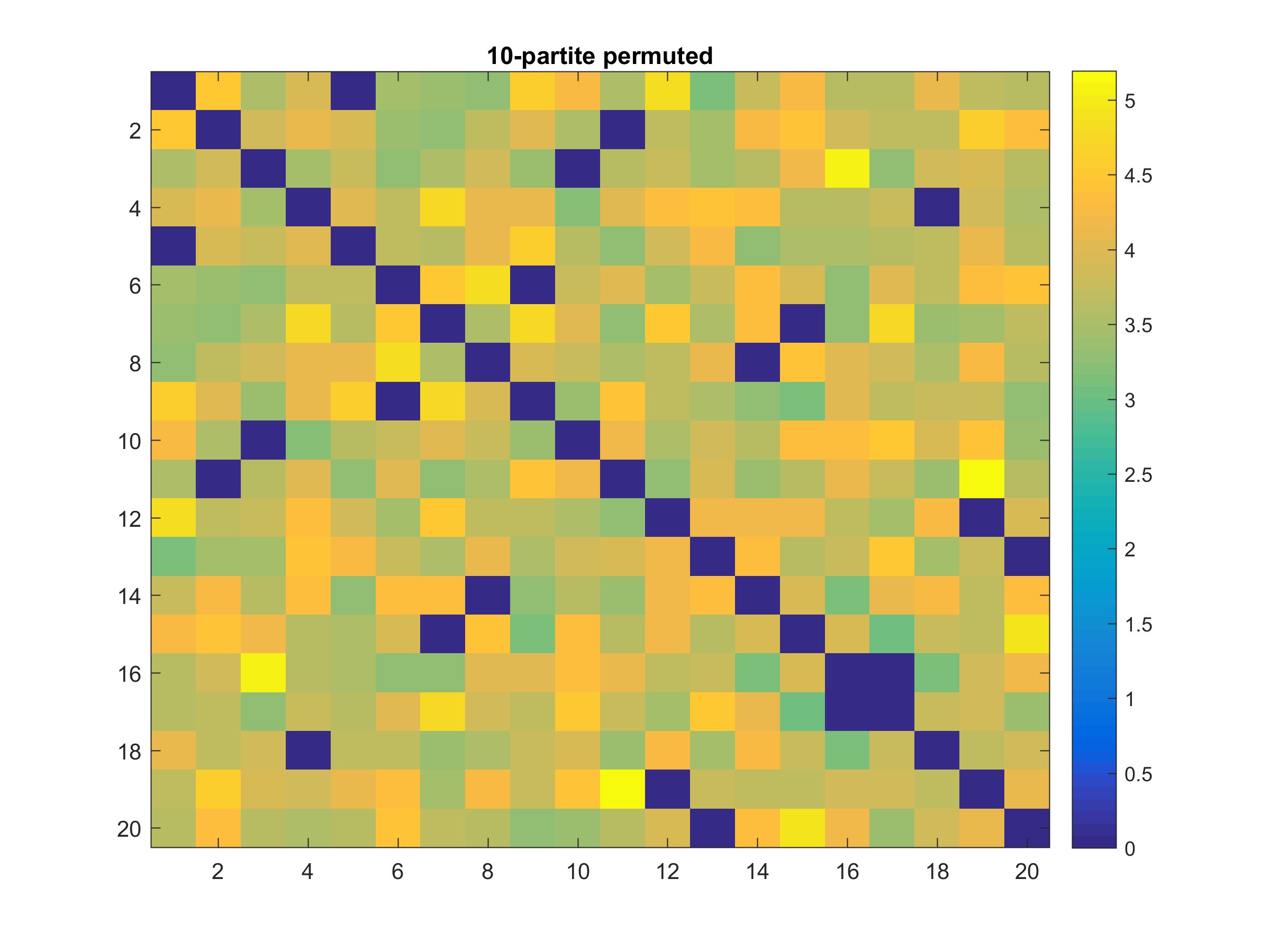}\label{fig2:f5}} 
  \hfill
  \subfloat[Detected 10-partite graph by the KPGD algorithm]{\includegraphics[width=0.5\textwidth]{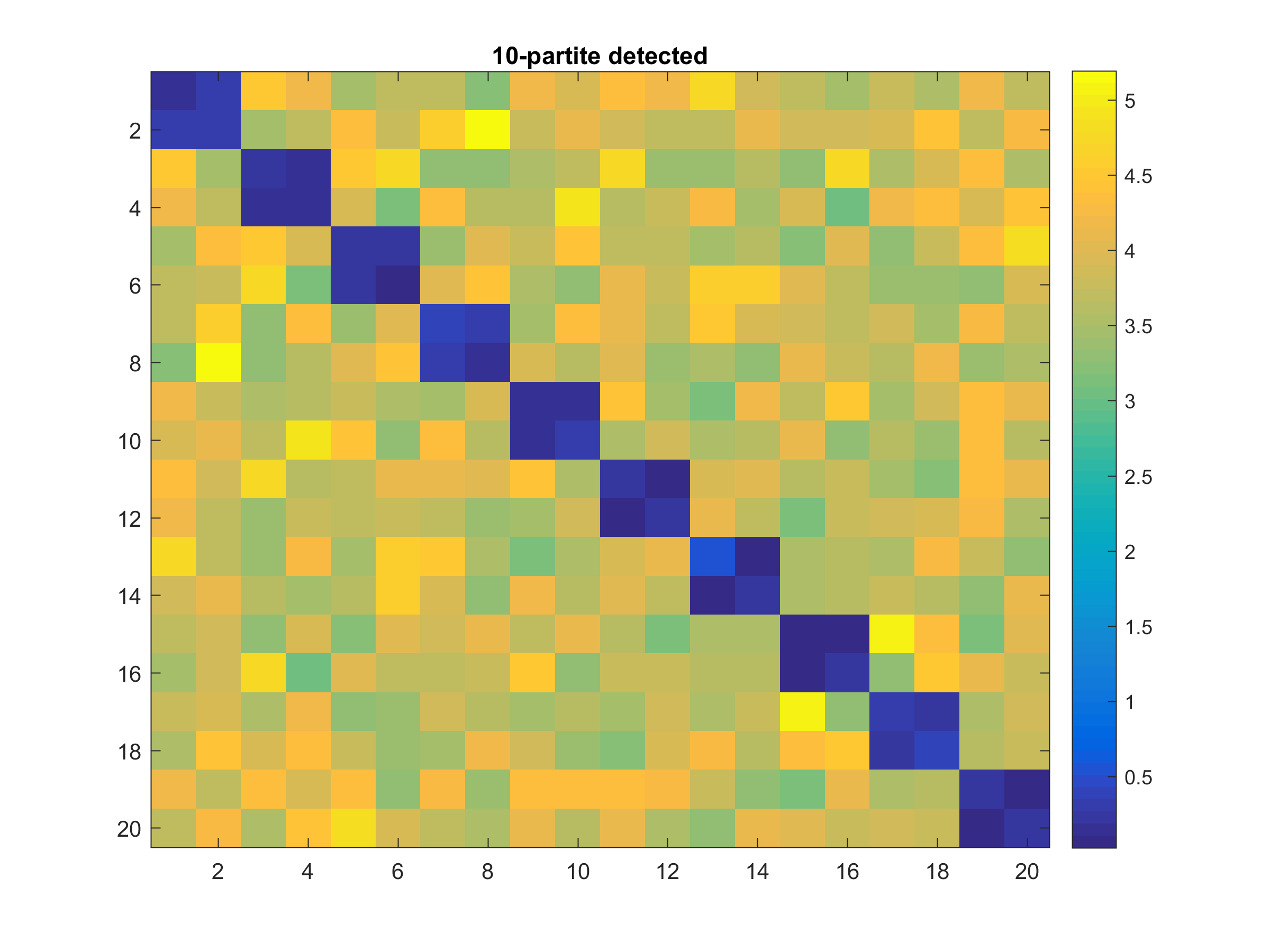}\label{fig2:f6}} 
  \caption{Applying our approach to simulated data sets: the left side figures are input $G_c$ with latent k-partite structure and the right side figures are the results of the KPGD algorithm with the apparent k-partite structure.}
\end{figure}

\begin{table*}[!ht]
\caption{Kpartite simulation}
 \label{L'WL and FDR}
 \begin{adjustbox}{width=1\textwidth}
 \small
 \begin{tabular}{c|rr|rr|rr|rr}
                                                                            \toprule
                                                                                Difference $\delta$ &\multicolumn{2}{c|}{$K=2$} & \multicolumn{2}{c|}{$K=4$} & \multicolumn{2}{c|}{$K=5$}& \multicolumn{2}{c}{$K=10$} \\
                                                                                & \multicolumn{1}{c}{$FP~mean(std)$} & \multicolumn{1}{c|}{$FN~mean(std)$} & \multicolumn{1}{c}{$FP~mean(std)$} & \multicolumn{1}{c|}{$FN~mean(std)$} &\multicolumn{1}{c}{$FP~mean(std)$} & \multicolumn{1}{c|}{$FN~mean(std)$}&\multicolumn{1}{c}{$FP~mean(std)$} & \multicolumn{1}{c}{$FN~mean(std)$}\\
                                                                                \hline
                                                                                3              &0.02(0.14)         &0.78(1.52)         &0.47(2.01)         &0.82(1.16)                &0.41(1.7)           &0.93(1.15)         &1.03(1.82)         &1.36(1.1)\\
                                                                                2.5 &0.08(0.94)         &0.75(1.59)         &0.29(1.52)         &0.92(1.2)                &0.6(1.97)           &1.03(1.28)         &1.21(1.85)         &1.5(1.24)\\
                                                                                3.5          &0.08(1.05)         &0.81(1.69)         &0.42(1.89)         &0.94(1.25)                &0.5(1.85)           &0.92(1.05)         &0.96(1.7)           &1.36(1.11)\\
                                                                                4              &0.06(0.71)         &0.58(1.32)         &0.22(1.41)         &0.88(1.11)                &0.46(1.92)         &0.9(1.15)           &1.07(1.87)         &1.32(1.21)\\
                                                                                4.5          &0.09(0.1)           &0.84(1.66)         &0.41(1.85)         &0.77(1.11)                &0.37(1.63)         &0.86(1.17)         &1.1(1.89)           &1.46(1.21)\\
                                                                                5              &0.16(1.5)           &0.91(1.68)         &0.45(1.91)         &0.86(1.2)                &0.32(1.54)         &0.88(1.12)         &0.89(1.84)         &1.46(1.13)\\
\bottomrule
  \end{tabular}
 \end{adjustbox}
 \end{table*}

\section{Data example}
We apply our method to an fMRI study for Parkinson's disease research (\citealp{Bowman16}). The fMRI data are acquired using a multi-slice ZSAGA sequence, yielding 30 axial slices (4mm thick) covering the entire cerebrum. Subjects lay supine with eyes open, maintaining attention to a visual fixation point on the computer screen, without other explicit tasks.  The data include 42 subjects, with 24 Parkinson’s patients (PD) patients and 18 healthy control subjects. The data were preprocessed in AFNI, including included slice-timing correction, co-registration, spatial normalization, and regional parcellation using AAL. We refined our rs-fMRI preprocessing to retain desired fluctuations associated with pial fluid, known to correlate with gray matter BOLD fluctuations, and to discard unwanted contributions from ventricular CSF and white matter, typically unrelated to neural activity. For the region level signal, we average the preprocessed time courses across all subjects for 90 AAL regions (\citealp{Tzourio02}).

We calculate 4005 Pearson correlation coefficients between the time courses of all pairs of 90 AAL regions.  We then perform two sample t-tests 
 on Fisher's Z transformed correlation coefficients and calculate  the whole graph matrix $\textbf{W}$, with the entry $i,j$ of $W_{ij}= -\log(p_{ij})$.  We apply the Pard algorithm for parsimonious differential brain connectivity network detection on $\textbf{W}$, which seeks to capture the most significantly differentially expressed connectivity edges within a smaller number of nodes.   We detect three networks  based on permutation tests of the Pard algorithm with all $P<0.001$. We further apply the KPGD algorithm on the three networks, and only one network is detected with the k-partite structure including 23 AAL regions such as orbito-frontal cortex, parietal region, basal ganglia, and limbic gyrus. 

\begin{figure}[!htp]
  \centering
  \subfloat[Input \textbf{W}]{\includegraphics[width=0.5\textwidth]{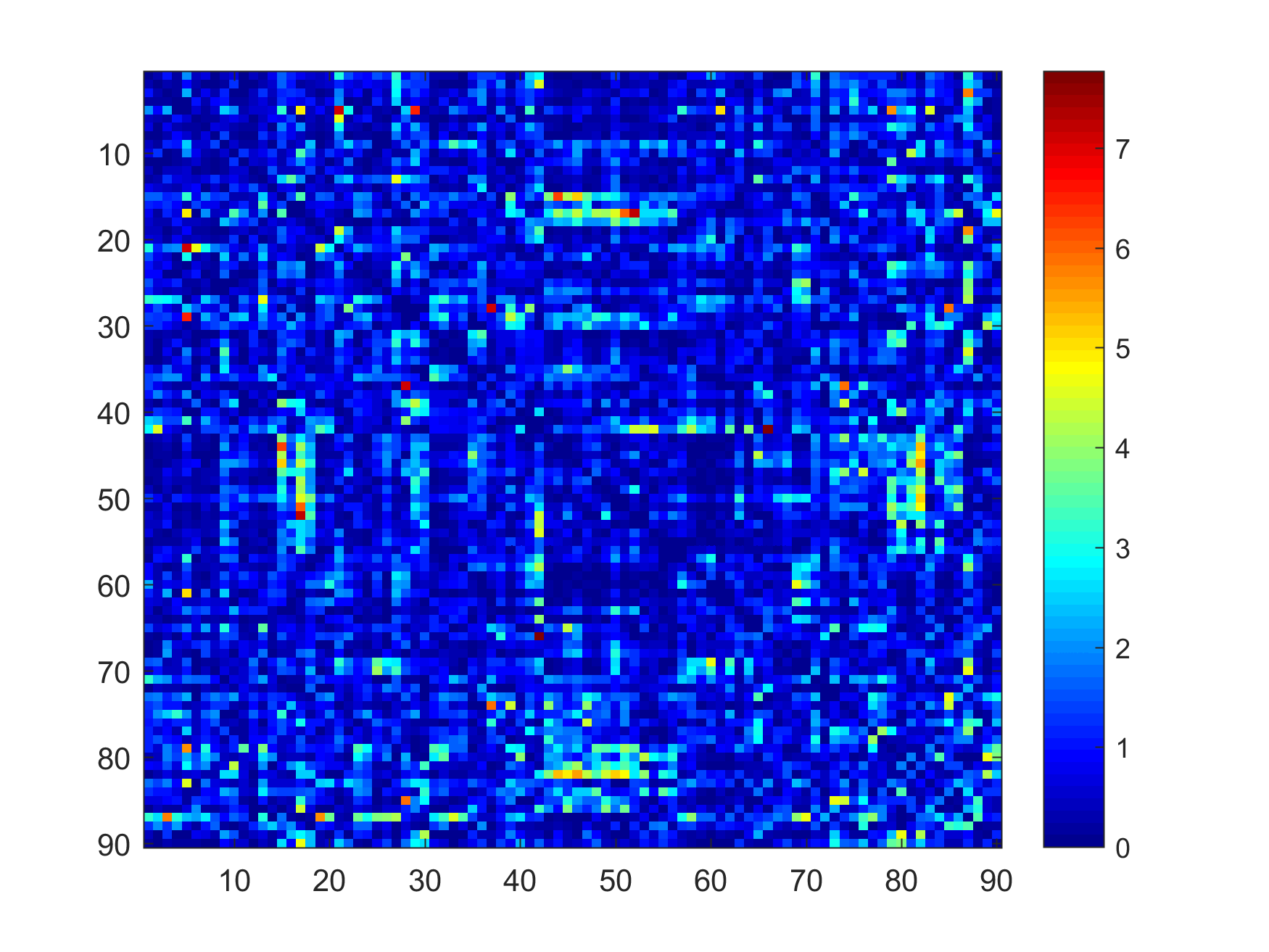}\label{fig3:f1}}
  \hfill
  \subfloat[Networks detected by the Pard algorithm]{\includegraphics[width=0.5\textwidth]{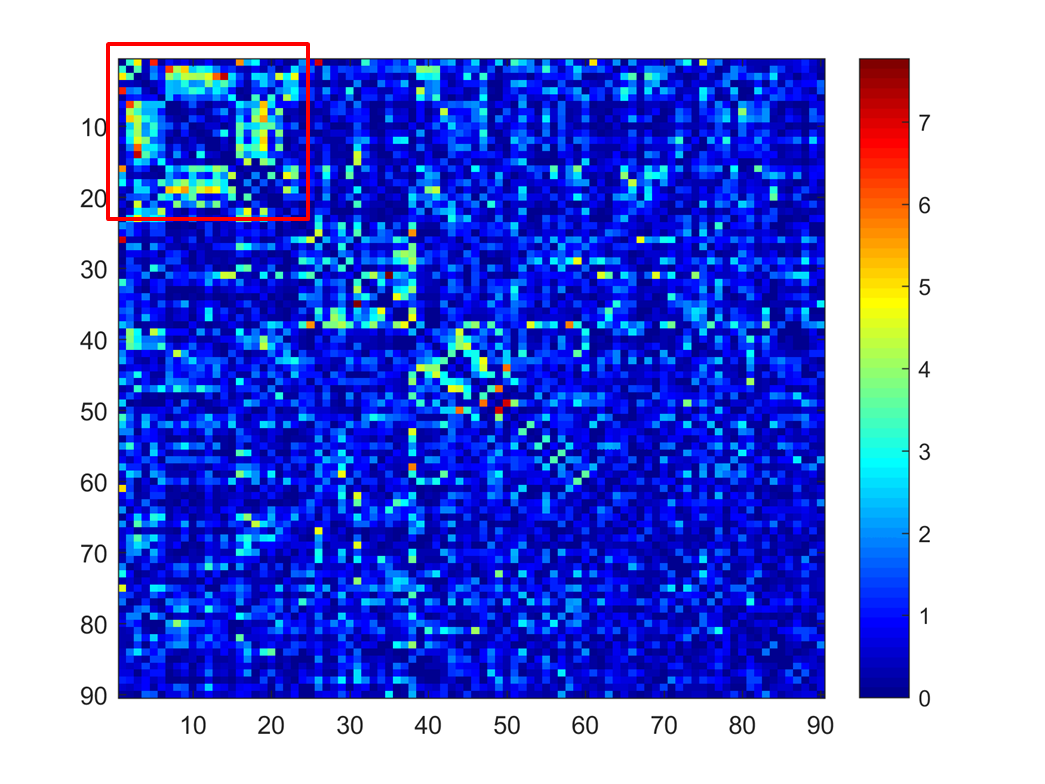}\label{fig3:f2}}
  \hfill
  \subfloat[After applying KPGD to (b): the reordered first network $\pi^{KPGD}(\textbf{W}_1)$ shows a k-partite graph topology   ]{\includegraphics[width=0.5\textwidth]{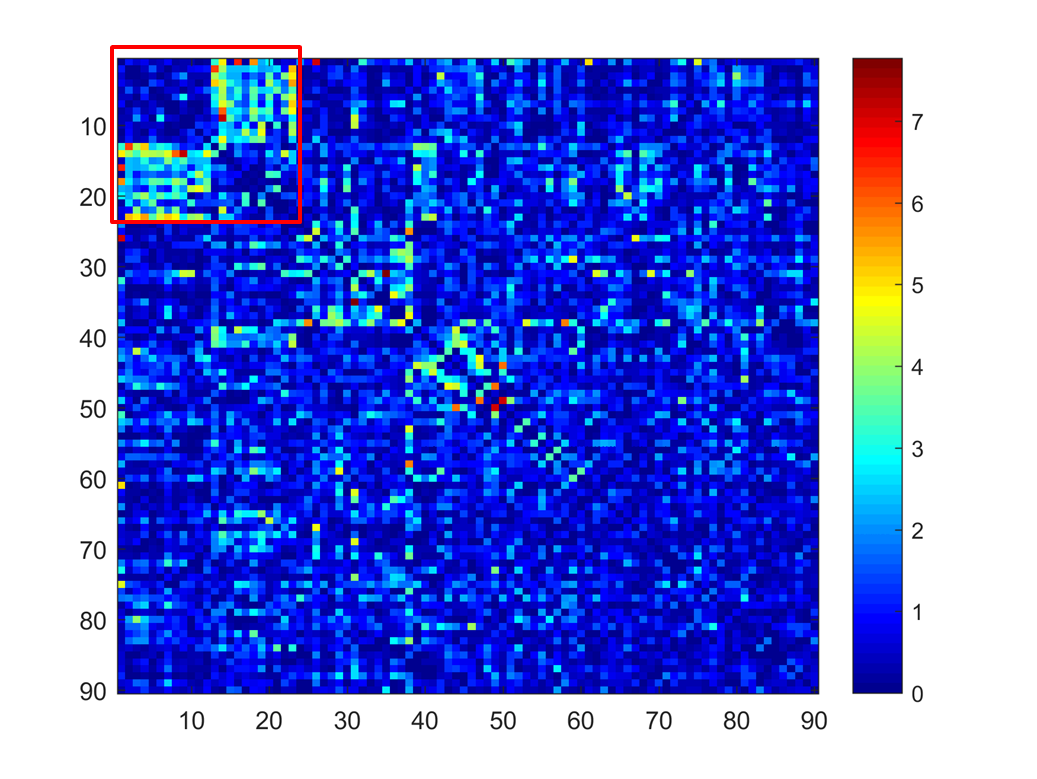}\label{fig3:f3}}
  \hfill
  \subfloat[GEP test of k-partite strcutre]{\includegraphics[width=0.5 \textwidth]{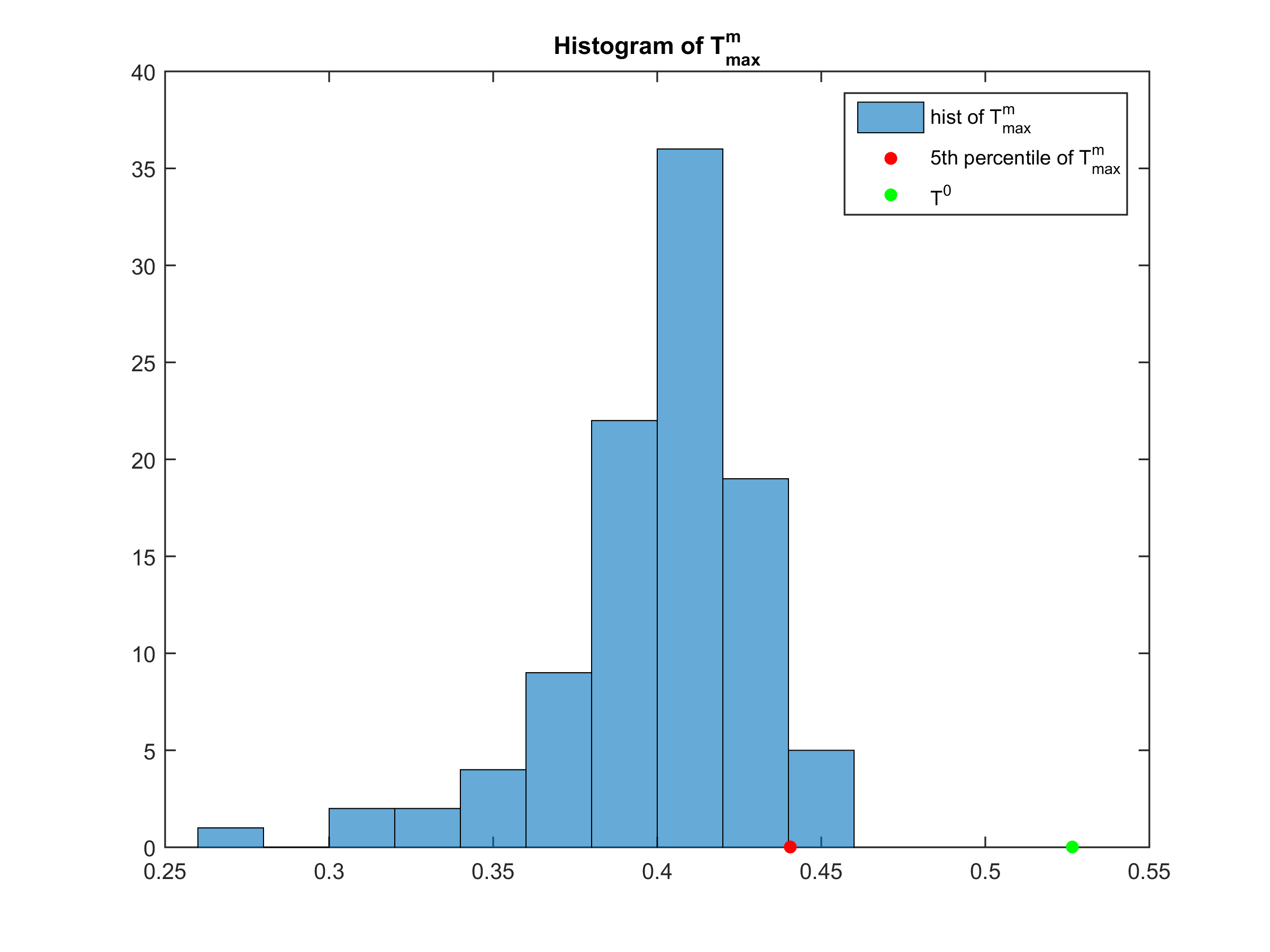}\label{fig3:f4}}
  \hfill
  \subfloat[Enlarged heatmap of $\pi^{KPGD}(\textbf{W}_1)$ with AAL region names]{\includegraphics[width=.5\textwidth]{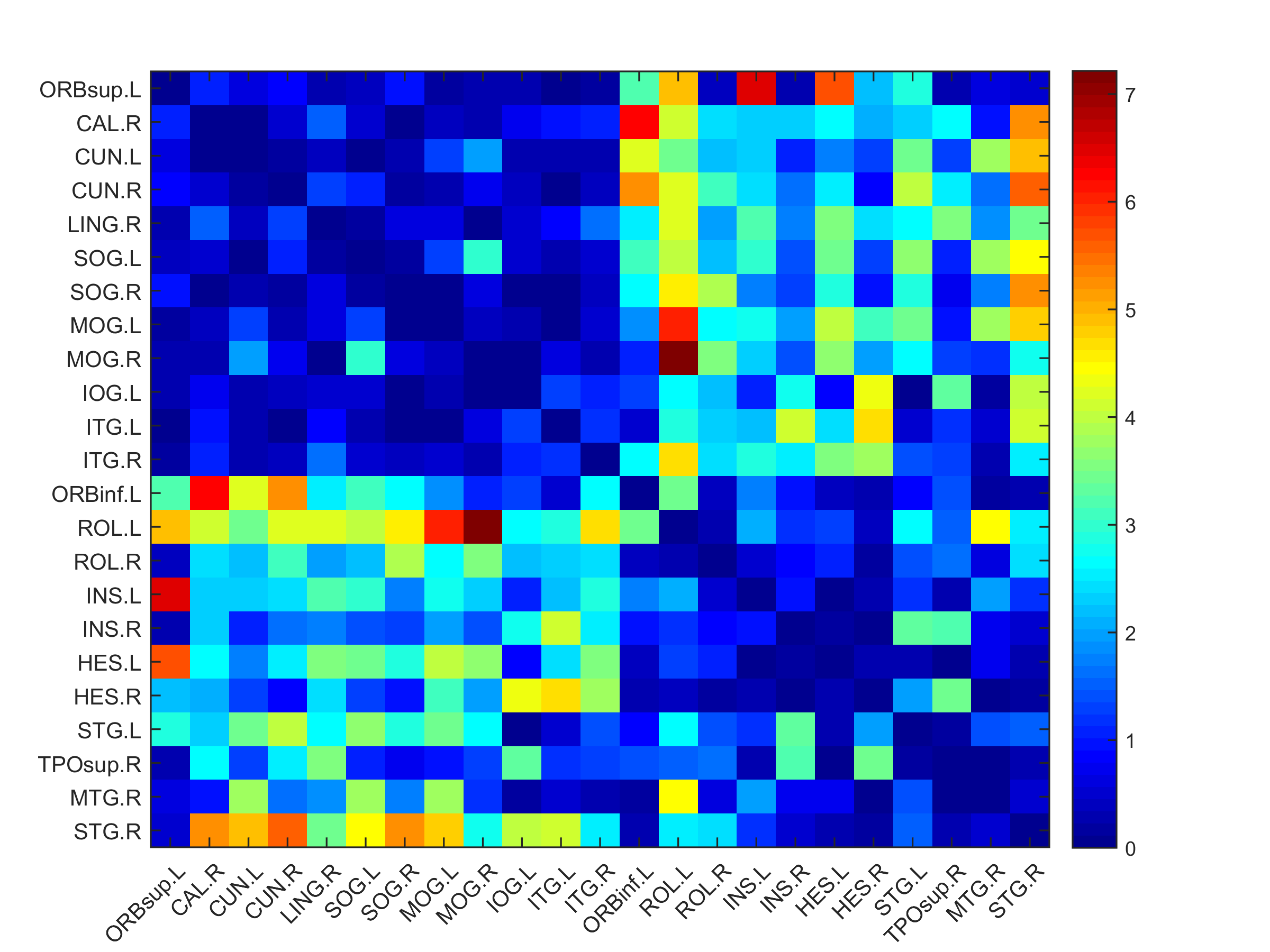}\label{fig3:f5}} 
  \hfill
\caption{Application of the NICE to the example data set: we first apply the Pard algorithm to obtain (b) from (a), and we apply our methods to all networks of (b) and find one subgraph with k-partite graph topology (c) and (e).}   
\end{figure}

The overall network detection procedure is demonstrated by Figure 3. Figure \ref{fig3:f1} displays our input data, which  is a $90 \times 90$ matrix of testing results $–\log(p_{ij})$, and the original distribution of differentially expressed edges (hot color) is also shown in the heatmap. Next, we apply the Pard algorithm to detect whether the informative edges are distributed in diagonal blocks. Figure \ref{fig3:f2} shows the detected parsimonious (small sized) networks that capture the most informative edges. We perform k-partite graph detection and statistical testing on each of the detected networks of the Pard algorithm. For example, we highlight (red circle) the first detected network by Pard in Figure \ref{fig3:f2}, and the resulting k-partite structure by applying our KPGD algorithm and GEP test on $\textbf{W}_1$ in Figure \ref{fig3:f3}. Based on the GEP testing results, only the first network's k-partite structure is significant (with $p<0.001$). The permutation procedure and test statistic appear in Figure \ref{fig3:f4}. 

We enlarge the detected k-partite network in Figure \ref{fig3:f5}. There are two large independent sets: set one mainly includes insular cortices, occipital lobes and frontal lobes, and set two mainly includes central frontal lobes and temporal lobes (see table \ref{tb2}). In set one, the altered connectivity of regions from the occipital lobe of the patients with PD which are well documented in \citealp{Occ1} and \citealp{Occ2}. Similarly, the  functions of insular cortex are linked with many symptoms of the Parkinson's disease (\citealp{Ins3}; \citealp{Ins2}; and \citealp{Ins1}). The findings regarding temporal lobes in set two have also been identified in previous studies (e.g. \citealp{Temp1}; \citealp{Temp2}). Overall, most differentially expressed edges  in the detected k-partite subgraph based on data analysis coincide with findings of numerous precedent studies (\citealp{Temp3}).

\begin{figure}[!htp]
  \caption{3D demonstration of the differentially expressed connectivity network with k-partite graph topology: red nodes are brain regions from set one and blue nodes for set two; yellow edges indicate controls $>$ cases and green edges indicate controls $<$ cases, the width of an edge represent the difference between the two groups. The normal control group shows hyper-connections for most differentially expressed edges in the k-partite structure.}
  \centering
    \includegraphics[width=0.7\textwidth]{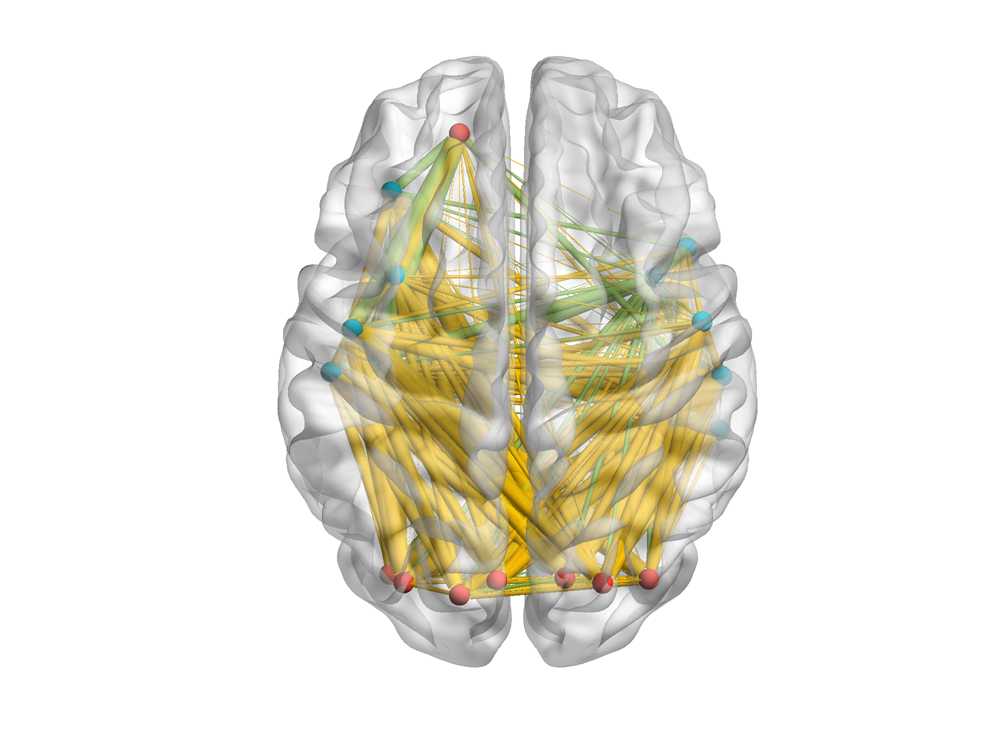}
    \includegraphics[width=0.7\textwidth]{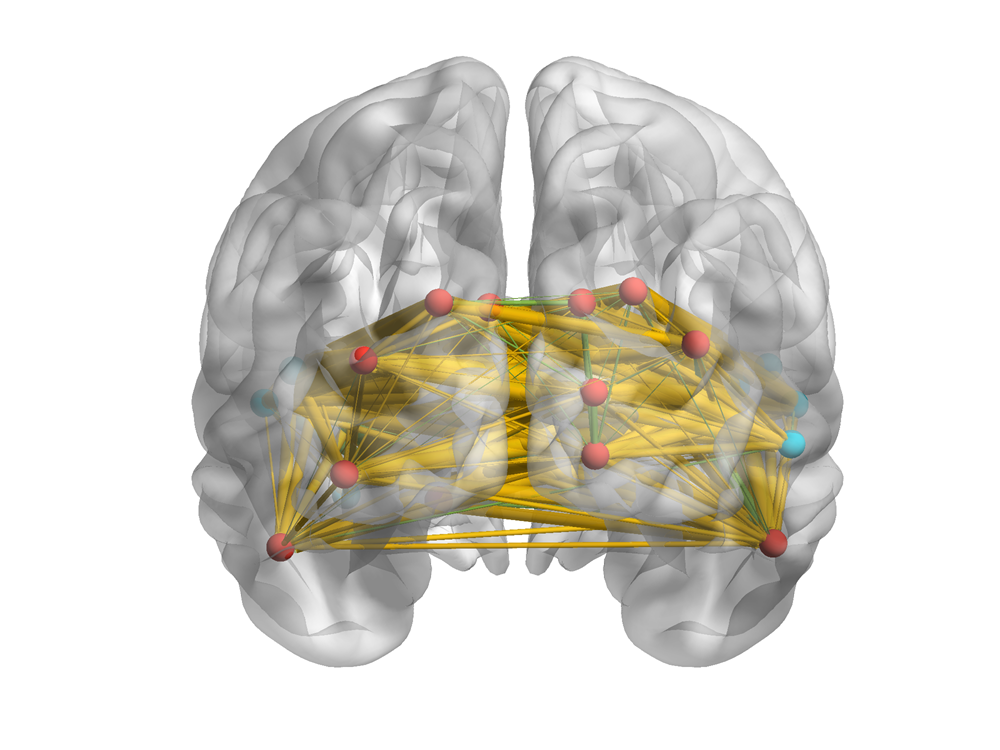}
    \label{3d1}
\end{figure}

In this article, we identify a differentially expressed connectivity network   with a latent bipartite topological structure. The network is not predefined but rather is automatically detected by our proposed methods.
In the detected differential network with a bipartite topological structure, most connections within each independent set are high for most of subjects, yet they have no significant difference in connectivity between normal controls and PD patients. Nevertheless, the connections between the two sets are differentially expressed between the two groups. We illustrate the k-partite structure in a 3D brain image (Figure \ref{3d1}). We note that The normal control group shows hyper-connections for most differentially expressed edges (around 85\%) in the detected k-partite subgraph (edge color code yellow in Figure \ref{3d1}). These results concur with the fact that Parkinson's disease  is a neurodegenerative disorder.  In addition,  the normal control group exhibits hyper-connections for most long-range differentially expressed edges such as edges from occipital lobes and inferior temporal lobes to insular and superior temporal lobes. There is only a small proportion of the diffrentially expressed edges that the PD patients express  hyper-connections than the normal controls, mainly including edges connected with the nodes of insular(R) or superior frontal gyrus orbital part. 
The structure apparent in our detected ``k-partite phenomenon"  may reflect  neuropathology of the Parkinson's disease.


\begin{table}[]
\centering
\caption{AAL regions in the k-partite graph}
\label{tb2}
\begin{tabular}{lllc|c|c|c}
\toprule
AAL region ame                                                                              & abbrevation &  index & x      & y      & z      & Set \\
\hline
\begin{tabular}[c]{@{}l@{}}Superior frontal gyrus, orbital part,\\   Left\end{tabular}      & ORBsup.L    & 5             & -17 & 47   & -13  & 1   \\

\begin{tabular}[c]{@{}l@{}}Inferior frontal gyrus, orbital part,\\   Left\end{tabular}      & ORBinf.L    & 15            & -36 & 30   & -12  & 2   \\
Rolandic operculum, Left                                                                    & ROL.L       & 17            & -47  & -8   & 14  & 2   \\
Rolandic operculum, Right                                                                   & ROL.R       & 18            & 53  & -6   & 15  & 2   \\
Insula, Left                                                                                & INS.L       & 29            & -35  & 7   & 3    & 2   \\
Insula, Right                                                                               & INS.R       & 30            & 39  & 6    & 2    & 2   \\
\begin{tabular}[c]{@{}l@{}}Calcarine fissure and surrounding cortex,\\   Right\end{tabular} & CAL.R       & 44            & 16  & -73  & 9    & 1   \\
Cuneus, Left                                                                                & CUN.L       & 45            & -6  & -80  & 27   & 1   \\
Cuneus, Right                                                                               & CUN.R       & 46            & 14  & -79  & 28   & 1   \\
Lingual gyrus, Right                                                                        & LING.R      & 48            & 16  & -67 & -4  & 1   \\
Superior occipital gyrus, Left                                                              & SOG.L       & 49            & -17 & -84 & 28  & 1   \\
Superior occipital gyrus, Right                                                             & SOG.R       & 50            & 24  & -81 & 31  & 1   \\
Middle occipital gyrus, Left                                                                & MOG.L       & 51            & -32 & -81 & 16   & 1   \\
Middle occipital gyrus, Right                                                               & MOG.R       & 52            & 38  & -80  & 19   & 1   \\
Inferior occipital gyrus, Left                                                              & IOG.L       & 53            & -36  & -78  & -8  & 1   \\
Heschl gyrus, Left                                                                          & HES.L       & 79            & -42 & -19 & 10   & 1   \\
Heschl gyrus, Right                                                                         & HES.R       & 80            & 46  & -17  & 10   & 2   \\
Superior temporal gyrus, Left                                                               & STG.L       & 81            & -53  & -21 & 7    & 2   \\
Superior temporal gyrus, Right                                                              & STG.R       & 82            & 58   & -22 & 7    & 2   \\
\begin{tabular}[c]{@{}l@{}}Temporal pole: superior temporal gyrus,\\   Right\end{tabular}   & TPOsup.R    & 84            & 48   & 15  & -17 & 2   \\
Middle temporal gyrus, Right                                                                & MTG.R       & 86            & 57   & -37  & -1   & 2   \\
Inferior temporal gyrus, Left                                                               & ITG.L       & 89            & -50 & -28  & -23 & 1   \\
Inferior temporal gyrus, Right                                                              & ITG.R       & 90            & 54  & -31  & -22  & 1  \\
\bottomrule
\end{tabular}
\end{table}

%

\newpage

\section{Discussion and Conclusion}

Functional integration has been a longstanding principle underlying complex human brain function, enabling 
interactions between different brain areas. From a network perspective, the interactions are assumed to be highly organized, like most intelligent networks, rather than distributed randomly. Neurological disorders, such as the Parkinson's disease, and psychiatric disorders may reveal patterns of dysregulation in such systems level characterizations of brain function.  Our work provides an illustration of how differentially expressed connectome features (between PD patient and normal controls) are also distributed in an organized topology. Our proposed methods provide a pathway to reveal the phenotype-related connectome features along with the latent topological structure and to conduct statistical tests about this organized topology.  Detected connectivity patterns may serve as useful features, e.g. as network biomarkers, in the diagnosis and treatment of brain disorders.

From an analytical point of view,   brain connectivity analyses  naturally involve with high dimensionality and graph topological properties because the connectivity features are interactions between brain areas. There has been an emergence of statistical techniques to handle high-dimensional data, for example, several multiple-testing strategies as well as shrinkage techniques including lasso, SCAD, and elastic network. Many of these methods are motivated by  high-throughput omics data where interest focuses on selecting   differentially expressed `nodes' (e.g. gene/protein expressions and brain activation). 

However, these techniques have not kept the pace to meet the needs of connectivity analyses, which consider edges reflecting associations between nodes. Complicating issues include the massive number of edges as well as the interdependence of edges in an organized and complex graph topological structure.   Often, the graph topology is highly informative for the population level analysis of  `edge' type features, and tailored statistical methods are  needed. Object oriented statistical analysis seems well-suited for the brain connectivity data sets (\citealp{Wang07}; \citealp{Marron14}), prompting the need for novel algorithms to recognize latent topological structures. Algorithms including NBS, Pard, and KPGD are developed for this purpose. Permutation tests are often used to test the statistical significance of the detected subgraphs. In this article, we have introduced a specific yet common network topological structure: k-partite graph topology along with the topology structure detection algorithm and corresponding statistical inference techniques.


From the perspective of neurophysiology and neuropathology, the detected graph topology structure not only improves the statistical power, but also could reveal important underlying disease mechanisms. For example, the Pard algorithm detects a general clique/complete subgraph structure that most edges in small network are differentially expressed among clinical subgroups. We develop more advanced tools to further examine  whether the detected clique/complete subgraph is isomorphic to a k-partite graph structure. The k-partite topological  reflects the organized distribution of brain phenotypes, which may uncover important characteristics of brain disorders, e.g. revealing long-range, rather than local, connections.

Although we use functional connectivity brain network data for demonstration, the ideas underlying our method are applicable to all types of connectivity data including functional connectivity (FC, e.g. from EEG and fMRI data) and structural connectivity (SC, e.g. from  diffusion-weighted imaging data) if the connectivity matrix represents an undirected graph. In addition, our method is also applicable for any choice of connectivity metrics, for instance, in FC analysis correlation coefficients, maximum information coefficient, or spectral coherence.


\section*{Acknowledgements}
Dr. Bowman's research work was funded by a grant from the NINDS (U18 NS082143) at NIH as part of the Parkinson’s Disease Biomarker Program.

\end{document}